\documentclass{PoS}

\usepackage{amsmath}
\usepackage{amsfonts}

\newcommand{\R}{\mathbb{R}}
\newcommand{\C}{\mathbb{C}}

\newcommand{\sign}{\mathrm{sign}}

\newcommand{\dP}{\mathrm{d}P}
\newcommand{\dU}{\mathrm{d}U}

\newcommand{\dphi}{\mathrm{d}\phi}

\newcommand{\UOne}{\mathrm{U}(1)}

\title{Simulating gauge theories on Lefschetz thimbles}

\ShortTitle{Simulating gauge theories on Lefschetz thimbles}
\author{{Jan M. Pawlowski}\\ 
	Institut f\"ur Theoretische Physik, Universit\"at Heidelberg, \\ Philosophenweg 16, D-69120 Heidelberg, Germany.\\
	E-mail: \email{j.pawlowski@thphys.uni-heidelberg.de}}
\author{{Manuel Scherzer}\\
	Institut f\"ur Theoretische Physik, Universit\"at Heidelberg, \\ Philosophenweg 16, D-69120 Heidelberg, Germany.\\
	E-mail: \email{scherzer@thphys.uni-heidelberg.de}}
\author{{Christian Schmidt}\\ 
	Fakult\"at f\"ur Physik, Universit\"at Bielefeld, Postfach 100131, D-33501 Bielefeld, Germany.\\
	E-mail: \email{schmidt@physik.uni-bielefeld.de}}
	\author{{Felix P.G.~Ziegler}\\
	Institut for Matematik og Datalogi, Syddansk Universitet, \\Campusvej 55, 5230 Odense, Danmark.\\
        E-mail: \email{ziegler@imada.sdu.dk}}
\author{\speaker{Felix Ziesch\'e}\thanks{The authors C. Schmidt and F. Ziesch\'e acknowledge support by the Deutsche Forschungsgemeinschaft (DFG, German Research Foundation) through the CRC-TR 211
'Strong-interaction matter under extreme conditions' project number 315477589 TRR 211. This work is further supported by EMMI, the BMBF grant 05P18VHFCA and is part of and supported by the DFG Collaborative Research Centre SFB 1225 (ISOQUANT)
as well as by DFG under Germany's Excellence Strategy EXC-2181/1-390900948 (the Heidelberg Excellence Cluster STRUCTURES).
M.~Scherzer acknowledges support from DFG under grant STA 283/16-2. 
F.~P.~G.~Ziegler is supported by Heidelberg University. 
Additionally we thank Andrei Alexandru, I.O. Stamatescu and Alexander Lindemeier for helpful discussions.}\\
        Fakult\"at f\"ur Physik, Universit\"at Bielefeld, Postfach 100131, D-33501 Bielefeld, Germany.\\
        E-mail: \email{fziesche@physik.uni-bielefeld.de}}

\abstract{Lefschetz thimbles have been proposed recently as a 
possible solution to the complex action problem (sign problem) in
Monte Carlo simulations. Here we discuss pure abelian gauge theory
with a complex
coupling $\beta$ and apply the concept of Generalized
Lefschetz thimbles. We propose to simulate the theory
on the union of the tangential manifolds to the thimbles.
We construct a local Metropolis-type
algorithm, that is constrained to a specific tangential manifold.
We also discuss
how, starting from this result, successive
subleading tangential manifolds can be taken into account via a
reweighting approach. We demonstrate
the algorithm on $U(1)$ gauge theory in 1+1 dimensions and
investigate the residual sign problem.}

\FullConference{37th International Symposium on Lattice Field Theory - Lattice2019\\
		16-22 June 2019\\
		Wuhan, China}
\begin{document}

\section{Generalized Lefschetz thimbles}
The numerical sign problem plagues many theories from being
simulated at certain parameters with conventional Monte-Carlo 
techniques \cite{deForcrand:2010ys}. Since
this problem is representation dependent 
\cite{Gattringer:2016kco}, there are different ways
to alleviate or even eliminate it such as dual representations, specialised
Monte-Carlo techniques (density of states) or methods based on
extending the configuration space (Complex Langevin \cite{Aarts:2008rr}, Lefschetz thimbles \cite{Cristoforetti:2012su}).
If we want to calculate
a multi-dimensional integral
\[
\int\dphi~e^{-S[\phi]}\mathcal{O}(\phi)
\]
with compact real fields $\phi$, where the integrand is holomorphic,
we can complexify the fields $\phi$ and according to Cauchy's
theorem choose a submanifold in complexified space homotopic to real subspace 
to get the same result. Dealing with the sign problem means
in this case choosing a submanifold, where the fluctuations
of the phase of the integrand are reduced (see e.g.\cite{Mori:2017pne}).
Since the phase itself is non-holomorphic, it depends on the integration
manifold. In other words, the sign problem is representation dependent.\\
Lefschetz thimbles
are originally a basis of homology classes for complex varieties.
In our case, their representatives can be chosen to keep
the phase of our integrand (or just $e^{-iS_I}$) constant. Being a
basis, one can build a submanifold homotopic to the
original integration space from these.\\
A Lefschetz thimble is generally defined to be the union of flowlines
generated by the steepest descent equation of the action $S$
\begin{equation}
\frac{\mathrm{d}\phi}{\mathrm{d}t} = -\left(\frac{\delta S}{\delta \phi}\right)^*
\label{SteepestDescent}
\end{equation}
which end in a non-degenerate critical point $\phi_{\sigma}$. Since we are looking
at a gauge theory, every critical point is naturally degenerate and
the classical Picard-Lefschetz theory does not apply. But still we 
have the concept of Generalized Lefschetz thimbles \cite{Witten:2010cx},
which was also outlined for QCD \cite{Cristoforetti:2012su}:
Instead of critical points, we have seperate critical manifolds
spanned by the zero modes of the action. Complementary to the zero modes
on the critical manifold are the Takagi and Anti-Takagi modes, which
classically span the tangent spaces of the thimble and the anti-thimble,
if there are no zero modes. In the case of compact gauge groups,
one can choose a compact submanifold of the critical manifold,
whose dimension plus the number of the Takagis gives the real
dimension of the original integration space. If the degeneracy
comes from the gauge degrees of freedom, this can be spanned
by the real gauge transformations and is called \emph{gauge orbit}.
In the following, we will use this freedom to construct a local
update algorithm on the tangent space.

\section{(1+1)d-U(1) Lattice Yang-Mills theory}
\begin{figure}[t]
\includegraphics[height=5cm]{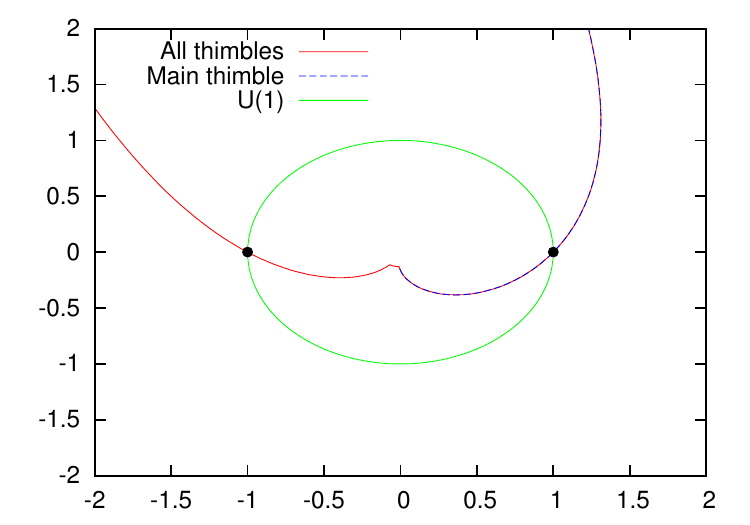}
\includegraphics[height=5cm]{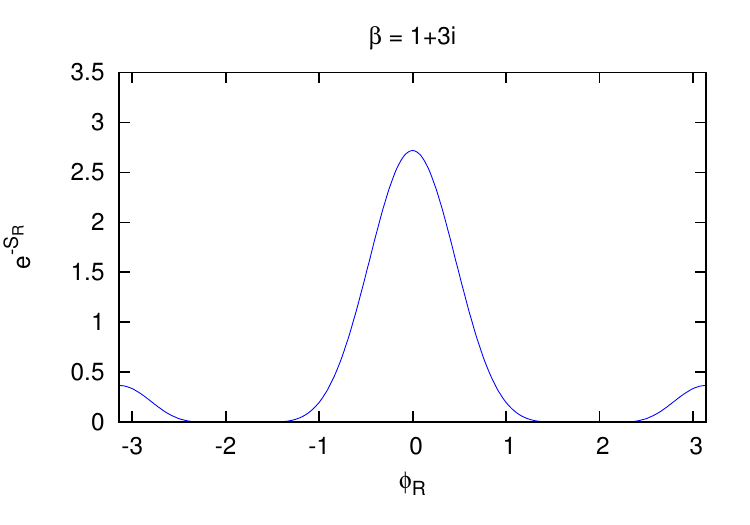}
\caption{This is the thimble manifold for $\beta = (1+3i)$ and their corresponding
plot of $e^{-S_R}$, which is proportional to the probability density.}
\label{PlotThimblesAndProbs}
\end{figure}

We discretize the Yang-Mills action on
a two dimensional
Euclidean space time lattice $\Lambda$, 
{\it i.e.} we consider Wilson's plaquette action
\cite{Wilson:1974sk}
\begin{equation}
  S = \beta\sum_{x}\left\{1-\frac{1}{2}
  \left(P_{01}(x)+P^{-1}_{01}(x)\right)\right\},
\label{wilson_action}
\end{equation}
where
$P_{01}(x)=U_0(x)U_1(x+\hat{0})U^{-1}_0(x+\hat{1})U^{-1}_1(x)$
denotes the elementary plaquette in the ($0,1$)-plane at site
$x$. The link variables $U_\mu(x)$ are elements of
the gauge group, which we consider to be $\text{U}(1)$.
Luckily, we have a formal solution for the partition sum
\begin{equation}
Z = \int\dU \exp\left({-S[U]}\right) = 
\sum_{n=-\infty}^{+\infty}\left[I_n(\beta)\right]^V
\label{MBesselExp}
\end{equation}
being a series in modified Bessel functions $I_n(\beta)$, where $V$ is the
number of plaquettes \cite{Balian:1974ts, Migdal:1975zg}. 
The sign problem is introduced by 
generalising to complex couplings $\beta$. This corresponds physically
to interpolating between imaginary and real time: Imaginary $\beta$ corresponds
to the real-time case, while real $\beta$ is the imaginary time case. In principal this 
allows to study thermal physics using paths in the complex time plane which for example approximate the Schwinger-Keldysh contour
\cite{Berges:2006xc, Makino:2015ooa}.
The critical manifolds obtained by setting the gradient of
the action to zero can be described by the
following relations
\begin{equation}
P_{01}(x) = P_{01}(x - \hat{\mu})\:\: \text{or}\:\: P_{01}(x) = -P^{-1}_{01}(x - \hat{\mu}),\;\;\mu\in\{0,1\}
\label{CritUN}
\end{equation}
between neighboring
plaquettes variables.
Additionally, we get the constraint
\begin{equation}
\prod_{x} P_{01}(x) = 1
\label{PerCon}
\end{equation}
by periodic boundary conditions.
This still leaves us a great amount of critical manifolds, since e.g. the constraint
gives us the possibility for every selection of $V$-th roots of
unity. We have to take a good selection of critical manifolds, in that
sense that the overall manifold, we are going to create is homotopic to
$[\UOne]^{2V}$ modulo copies. For inspiration, we look at the action for
one plaquette 
\begin{equation}
S = -\beta/2(P+P^{-1}),
\label{OnePlaqAct}
\end{equation}
where we omitted a volume factor.
We have naturally two critical points $P=\pm1$ with the respective
imaginary parts of the action being $S_I=\mp\beta_I$
and their attached thimbles $\mathcal{J}_0$, $\mathcal{J}_1$
, see Figure \ref{PlotThimblesAndProbs}.
We can take these possibilities for values of $P_{01}(x)$. Then we
are restricted by (\ref{PerCon}) to configurations, where an
even number of plaquettes can be $-1$. These configurations clearly
fulfill (\ref{CritUN}).

\subsection{Critical manifolds, their tangent spaces and local updates}
\begin{figure}[t]
    \centering
    \includegraphics[height=3.7cm]{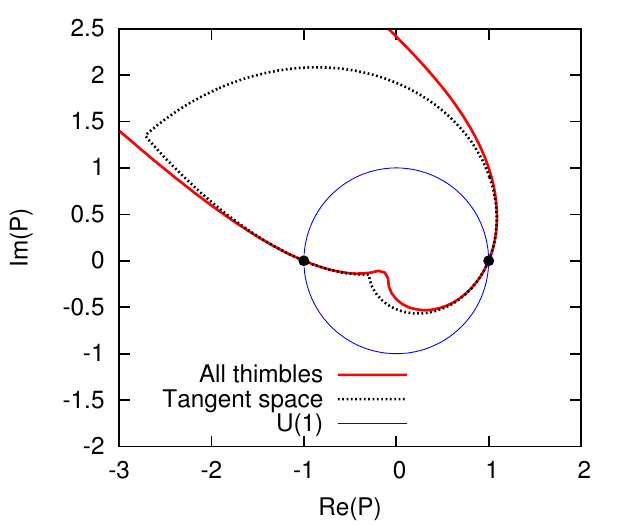}
    \includegraphics[height=3.7cm]{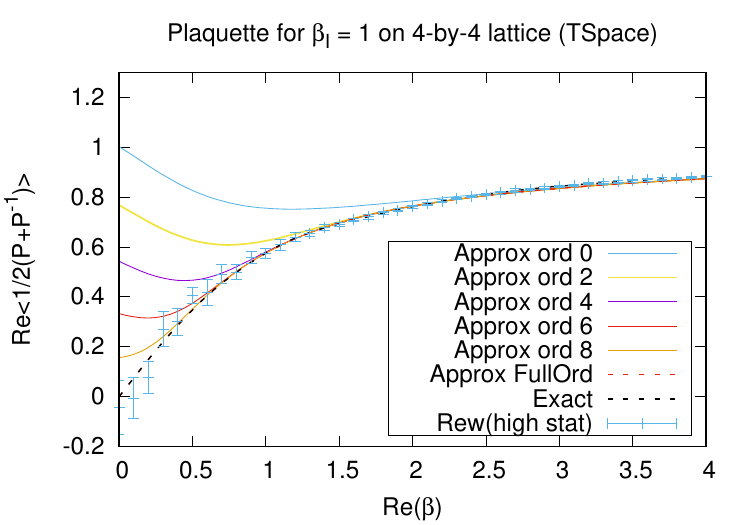}
    \includegraphics[height=3.7cm]{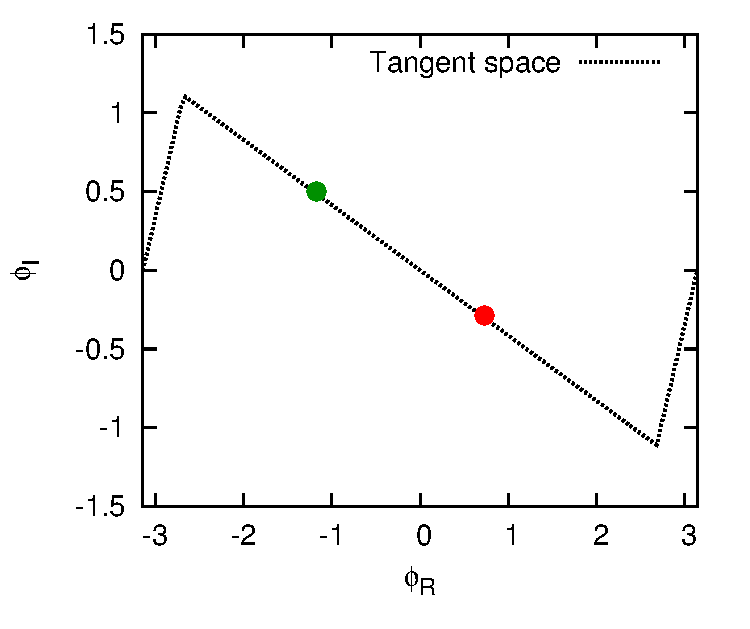}
    \caption{Left: Comparison of the thimbles and their tangent spaces for
    $\beta = 3+3i$. Middle: Thimble hierarchy depending on $\beta_R$ in the approximation.
    Right: Configurations on the tangent space of two plaquettes related to one link.}
    \label{FigApproxHierarchy}
\end{figure}

These are critical manifolds, since our degrees of freedom are still
links. We have to compute their Takagi and Anti-Takagis, i.e.
vectors, which are solution to the equation
\begin{equation}
\label{takagi}
H^*\Delta z^* = \lambda \ \Delta z\:\:\text{with}\:\:\lambda\in\R,
\end{equation}
where $H$ is the Hessian of the action. Positive $\lambda$ refer to
thimble directions, negative $\lambda$ to anti-thimble directions and
$\lambda = 0$ refers to zero-modes, which come e.g. from the gauge degrees
of freedom (see e.g. \cite{Cristoforetti:2012su,Alexandru:2015xva}). 
For our selection of critical manifolds $P_i=\pm 1$ the Hessian splits
into a real matrix with a complex prefactor $H=\beta M$, whose eigenvectors $v$
and eigenvalues $\alpha$ can be computed. For $\alpha\neq 0$ we get
for the Takagis
\begin{equation}
\label{Takagi}
\Delta z = \sqrt{\frac{\pm\sign(\alpha)\beta^*}{|\beta |}}v,
\end{equation}
where the sign denotes, if it is a Takagi or Anti-Takagi vector.\\
We observe, that our Hessian is independent from
the actual configuration in the critical manifold. Therefore,
we can deduce that the projection of the
subspace spanned by its zero modes in the Lie algebra is
the critical manifold itself.
\[
\left\{
U_\mu(x)^{\mathrm{crit}} = U_\mu(x)^{\mathrm{crit},0}
\exp\left({i\sum_{k=1}^{\#(\alpha=0)}c^kv^{x,\mu}_k(\alpha = 0)}\right)\;\;
| c^k\in\C,\;\mu\in\{0,1\},\; x\in\Lambda
\right\}
\]
Normally, we have to choose a gauge orbit
of real dimension by allowing only $c^k\in\R$, where we can span
the thimble using the Takagi directions.
For our main critical manifold $\{P_i=1\;\forall i\}$, the complex
prefactor in (\ref{Takagi}) is the same for all Takagi modes.
By tilting the real zero modes with the same factor, we differ
from a gauge orbit, but get a manifold, which still gives
the same expectation values for observables invariant 
under the zero modes. The set of tilted zero modes and Takagis is
equivalent to the tilted unit basis.
This gives us the possibility to have a local update
algorithm on the tangent space. This is naturally computationally far less
demanding than sampling on the thimbles themselves.
Restricting ourselves to the main
critical manifold, the Jacobian is constant and drops
out for expectation values. Another reason, why this
could be feasible is the closeness of the local
tangent space to the thimble in the one-plaquette model
(see Figure \ref{FigApproxHierarchy}) in their important
regions.
\subsection{Hierarchy of critical manifolds}
Our selection of critical manifolds has a
natural hierarchy, which is reflected by the values of
the action $S = 4k\beta,\:\:\: k=0,\ldots,\lfloor{\frac{V}{2}}\rfloor,
$ where $k$ is the number of turned plaquette pairs, which are $-1$. 
With increasing $\beta_R$ their weights in 
the partition sum
decrease exponentially. Note that on the thimble the real 
part of the action becomes minimal at the critical manifold. 
Consequently for pure imaginary $\beta$,
every thimble contributes equally. Otherwise we get
a close result by taking into account only a few thimbles, since the others
are exponentially suppressed. To get a hint on how strong this is the
case, we approximate our model by just taking the leading order
contribution of our formal expansion in modified Bessel functions
(\ref{MBesselExp})
\begin{equation}
Z = \left[\int_{\UOne}\dP \ e^{\beta/2(P+P^{-1})}\right]^V = \left[I_0(\beta)\right]^V.
\end{equation}
Physically this corresponds to removing periodic boundary conditions.
Since this is the One-plaquette model to the power of the volume,
we can expand this in term of its thimbles $\mathcal{J}_0$, $\mathcal{J}_1$
\begin{equation}
Z  =  \left[\int_{\mathcal{J}_0}\dP \ e^{\beta/2(P+P^{-1})} + \int_{\mathcal{J}_1}\dP \  e^{\beta/2(P+P^{-1})}\right]^V\\
 =:  \left[Z_0 + Z_1\right]^V
 = \sum_{k=0}^V
\begin{pmatrix}
V \\
k
\end{pmatrix}
Z_0^{V-k} \ Z_1^k.
\end{equation}
Using these, we can calculate approximate values for our observables
and their dependence on the thimble hierarchy (see Figure \ref{FigApproxHierarchy}).

\section{Simulation and comparison}

\subsection{Algorithm}
Since our critical manifolds and tangent spaces
depend on the values of the plaquettes, we use them
to confine the regions where we sample on them. As one
can observe in Figure \ref{FigApproxHierarchy} the two
tangent spaces for one plaquette intersect and we can
glue them together. This union is homotopic
to the orginal $\UOne$ group. We will use these intersections
to limit the region of tangent space we explore.
All in all we do a local Metropolis update on the links,
where we control the values of the associated plaquettes,
preventing them from wandering off the designated main tangent space shown 
on the right hand side of
Figure \ref{FigApproxHierarchy}. This is guaranteed 
by defining that candidates which would land beyond the edges have probability zero
and will be rejected by the Metropolis.

\subsection{Results for the main tangent space}
\begin{figure}[t]
\begin{center}
\includegraphics[height=4.2cm]{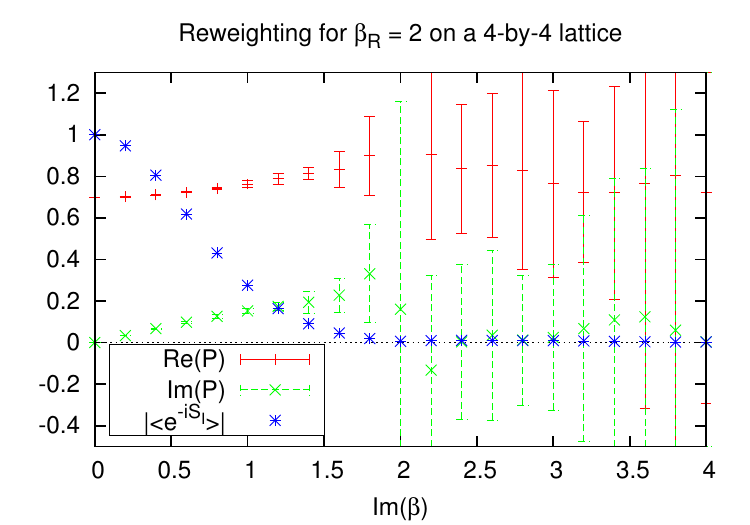}
\includegraphics[height=4.2cm]{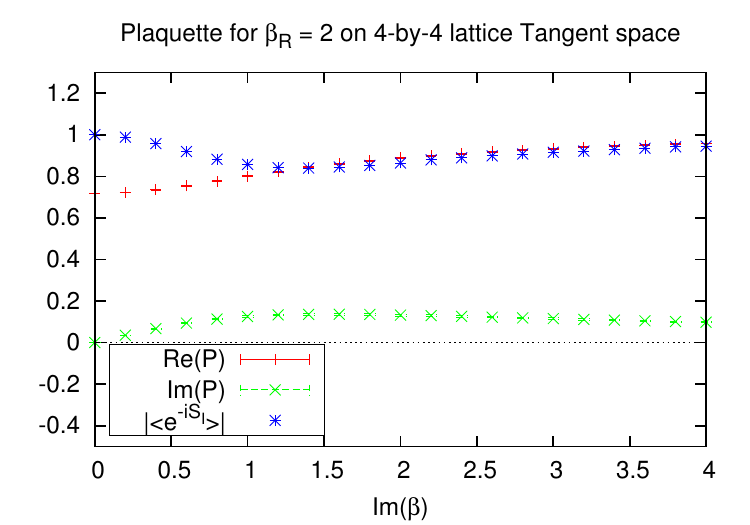}
\includegraphics[height=4.2cm]{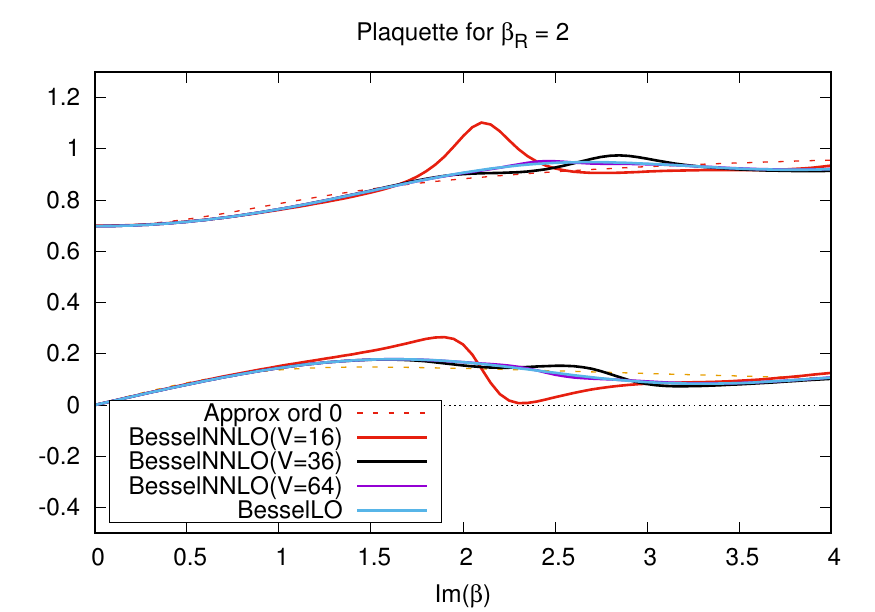}
\includegraphics[height=4.2cm]{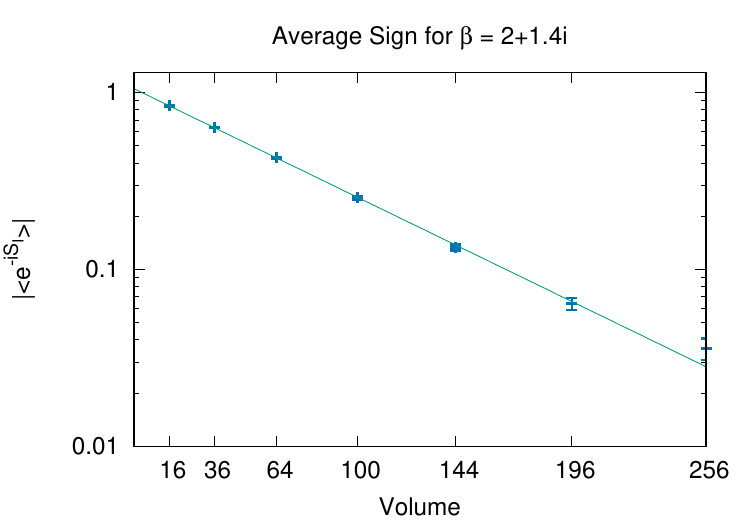}
\includegraphics[height=4.2cm]{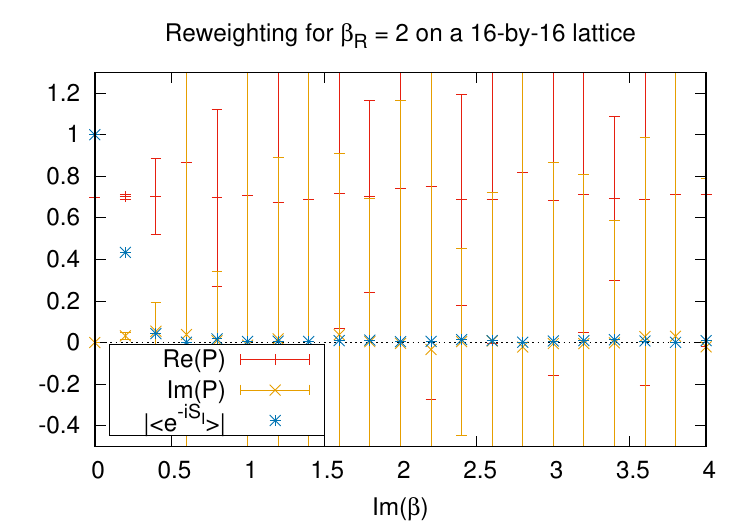}
\includegraphics[height=4.2cm]{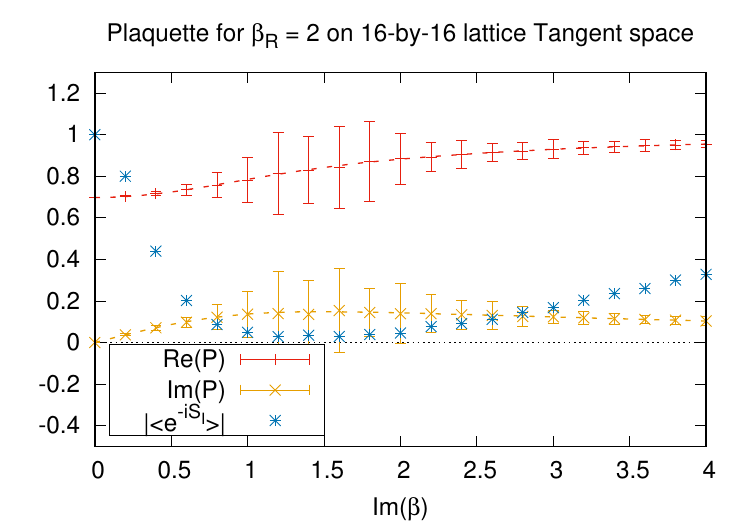}
\end{center}
\caption{
First row: plaquette for $\beta_R=2$ with reweighting compared to simulation on the main tangential manifold on a 4 x 4 lattice.
Second row: volume dependence of the formal solution and the average sign of the main tangential manifold.
Third row: plaquette for $\beta_R=2$ with reweighting compared to simulation on the main tangential manifold on a 16 x 16 lattice.
}
\end{figure}
We calculate the
expectation value of the average plaquette
$<1/2(P+P^{-1})>$. Since $\beta$ is complex,
this has a real and an imaginary part.
We first note that simulating on the main
tangent space alleviates the sign problem in comparison
with normal reweighting. Since we only take
the main tangent space into account, the results
can be considered 'right' only for large
enough beta and sufficiently large volumes. Another thing we note
is that for constant $\beta_R$, the average sign
$|<e^{-iS_I}>|$ has a minimum on the $\beta_I$
range for every observed volume. So it gets better again
for higher $\beta_I$.\\
Looking at this dip at $\beta = 2+1.4i$, we look
at the volume dependence of $|<e^{-iS_I}>|$ using
4-by-4 to 16-by-16 lattices. This seems strictly
exponential like it is predicted for the full
theory by considering the free energy. But even
for a 16-by-16 lattice, where reweighting fails very early
for small $\beta_I$, we can get a large range where
the sign problem is quite mild on the main tangent space.
Since higher orders in the modified Bessel function
expansion (\ref{MBesselExp}) are suppressed by
the volume, the leading order
approximation coincides stronger with the simulation result.

\subsection{Reweighting onto the other tangent spaces}
For taking another tangential manifold $\tau_1$ into account, 
e.g.~the one related to the critical manifold, where two
plaquettes are $-1$, which would be the next one in
the hierarchy, we need the ratio $Z_1/Z_0$
of the partition sums, since we have
\begin{equation}
<\mathcal{O}>_{\tau_1\cup\tau_2} =
\frac{\int_{\tau_1}\dU~\mathcal{O}[U]e^{-S[U]} + \int_{\tau_2}\dU~\mathcal{O}[U]e^{-S[U]}}
{\int_{\tau_1}\dU~e^{-S[U]} + \int_{\tau_2}\dU~e^{-S[U]}} =
\frac{<\mathcal{O}>_{\tau_0} + (Z_1/Z_0)<\mathcal{O}>_{\tau_2}}
{1+(Z_1/Z_0)}.
\end{equation}
Calculating this, we can follow the method proposed in \cite{Bluecher:2018sgj}:
Suppose, we have a mapping
\[
f\colon \tau_0  \longrightarrow\;\; \tau_1.
\]
Then we can write
\begin{equation}
\frac{Z_1}{Z_0} = \frac{\int_{\tau_0}\dU e^{-S[f(U)]+S[U]}\det[\mathrm{d}f]e^{-S[U]}}{\int_{\tau_0}\dU e^{-S[U]}}
 = <e^{-S\circ f+S}\det[\mathrm{d}f]>_0
\end{equation}
The problem here is to find a suitable $f$, which we discuss
in our upcoming paper \cite{UOneTSpace}. But we can already say, that
since we consider only tangent spaces, $f$ is linear and therefore
the Jacobian $\det[\mathrm{d}f]$ is be a constant factor.

\section{Summary and Outlook}
We have simulated a two dimensional U(1) lattice gauge theory on the tangent space of its
main thimble. Hereby, the sign problem is drastically reduced, while the computational
complexity has stayed the same as for a local Metropolis update. We proposed taking into
account subleading thimbles by a reweighting approach, which is under construction
and discussion. We will pursue the technique in the future by looking at other gauge groups
including also fermionic determinants and a chemical potential.

{\small
\bibliography{U1bib}
\bibliographystyle{apsrev4-1ow}
}
\end{document}